\documentclass{article}
\usepackage{graphicx} 
\usepackage{listings}
\usepackage{xcolor}
\usepackage{url}
\usepackage{authblk}
\usepackage{bookmark}
\usepackage{rotating}
\usepackage{lscape}
\usepackage{float}
\PassOptionsToPackage{table}{xcolor}
\usepackage[table]{xcolor}
\usepackage{longtable}
\usepackage[title]{appendix}
\usepackage[font=small,labelfont=bf]{caption}
\usepackage{adjustbox}
\usepackage{booktabs}
\usepackage{changepage}
\usepackage{array}
\usepackage{multirow}
\usepackage{MnSymbol}
\usepackage{placeins}
\usepackage{amsmath}
\usepackage{listings}
\usepackage{anysize}
\usepackage{hyperref}
\marginsize{15mm}{15mm}{15mm}{15mm}

\usepackage{longtable}

\lstdefinelanguage{json}{
    basicstyle=\ttfamily\footnotesize,
    numbers=none,
    numberstyle=\tiny\color{gray},
    numbersep=5pt,
    backgroundcolor=\color{lightgray!20},
    showspaces=false,
    showstringspaces=false,
    breaklines=true,
    frame=single,
    morestring=[b]",
    morecomment=[l]{//},
    moredelim=[s][keywordstyle]{:}{\ },
    keywordstyle=\color{blue}
}

\usepackage[backend=bibtex8,style=numeric,hyperref=true,backref=true,bibencoding=utf8,doi=true,url=true,maxcitenames=1]{biblatex}
\title{The Common Data Format (CDF): \\ A Standardized Format for Match-Data in Football (Soccer)}
\author[1]{Gabriel Anzer}
\author[2]{Kilian Arnsmeyer}
\author[2,3]{Pascal Bauer}
\author[4, 5, 13]{Joris Bekkers}
\author[6]{Ulf Brefeld}
\author[7]{Jesse Davis}
\author[8]{Nicolas Evans}
\author[9]{Matthias Kempe}
\author[8]{Samuel J Robertson}
\author[10, 11]{Joshua Wyatt Smith}
\author[7, 12]{Jan Van Haaren}

\affil[1]{RB Leipzig, Leipzig, Germany}
\affil[2]{Deutscher Fußball-Bund (DFB),
Frankfurt, Germany}
\affil[3]{Saarland University, Saarbrücken, Germany}
\affil[4]{U.S. Soccer Federation, Chicago, USA}
\affil[5]{UnravelSports, Breda, Netherlands}
\affil[6]{Leuphana University, L\"uneburg, Germany}
\affil[7]{KU Leuven, Leuven.AI, \& LISS, Heverlee, Belgium}
\affil[8]{FIFA, Zurich, Switzerland}
\affil[9]{University of Groningen, Groningen, Netherlands}
\affil[10]{Wyatt AI Inc., Montreal, Canada}
\affil[11]{Concordia University, Montreal, Canada}
\affil[12]{Club Brugge, Brugge, Belgium}
\affil[13]{PySport, Eindhoven, Netherlands}

\date{December 2024}

\begin{document}

\maketitle

\begin{abstract}

During football matches, a variety of different parties (e.g., companies) each collect (possibly overlapping) data about the match ranging from basic information (e.g., starting players) to detailed positional data. This data is provided to clubs, federations, and other organizations who are increasingly interested in leveraging this data 
to inform their decision making. Unfortunately, analyzing such data pose significant barriers because each provider may (1) collect different data, 
(2) use different specifications even within the same category of data, (3) represent the data differently, and (4) delivers the data in a different manner (e.g., file format, protocol).  Consequently, working with these data requires a significant investment of time and money.  The goal of this work is to propose a uniform and standardized format for football data called the \emph{Common Data Format (CDF)}. The CDF specifies a minimal schema for five types of match data: match sheet data, video footage, event data, tracking data, and match meta data. It aims to ensure that the provided data is clear, sufficiently contextualized (e.g., its provenance is clear), and complete such that it enables common downstream analysis tasks. 
Concretely, this paper will detail the technical specifications of the CDF, the representational choices that were made to help ensure the clarity of the provided data, and a concrete approach for delivering data in CDF. This represents Version 1.0.0 of the CDF. 




\end{abstract}

\section{Introduction}
Over the past two decades, professional sports organisations have increasingly relied on data-driven analysis to inform roster construction, tactical analysis, and marketing \cite{Link2018,Lolli_2024}. This change has arisen for several reasons. First, there have been notable public examples of teams successfully employing data,  which is most famously exemplified by baseball~\cite{Lewis2003}. Second, there has been a huge increase in the quantity and quality of data that can be collected, due in part to the growth of technology providers and associated financial incentives.   


In team invasion sports such as American football, basketball, and football (soccer) there typically six sources of data: 
\begin{description}
    \item[Match Sheet Data] include information such as the result of the match, line-ups, cards, and goals. Such data are collected for almost all matches, including amateur ones. 
    \item[Video Footage]  is recorded from multiple perspectives and in multiple resolutions for each match. These data serve various purposes such as providing broadcast to fans or facilitating tactical analyses. 
    \item[Event Data] record semantically meaningful “on-ball” events that occur during a match \cite{Pappalardo2019a} such as passes, tackles and shots.  Each event is annotated with information (e.g., location on the pitch, time-stamp, involved players). Events are typically collected manually by trained operators following a definition catalog. There are differences among vendors about the catalog of events and what information is recorded about each specific event. 
    \item[Tracking Data] record the positions (i.e., x,y,z-coordinates) of all players, the ball, and the referees multiple times per second~\cite{Manafifard2017a}. This data is typically collected using an in-venue camera system or extracted from broadcast video footage with the help of bespoke image recognition software.
    \item[Match Meta Data] contain contextual information such as crowd size, pitch dimensions, team colors, etc. 
    \item[Player Physical Data] describe measurements related to the players' physical constitution, e.g., heart rate data collected by wearable technologies. 
\end{description}
The most powerful and meaningful analyses often require combining the above types of data~\cite{Rein2016}. Specifically, this is important in areas such as monitoring workload, performing tactical analysis, and scouting players. 

Unfortunately, in practice, the way the data are collected and delivered pose a number of significant barriers that can prevent or limit performing such analyses. First, the definitions of key events are often ambiguous and vary across vendors, which is compounded by the fact that such events are often labeled by human annotators who make mistakes. Even in seemingly objective events such as shots, there are disagreements about how many of them occurred in a match~\cite{bialik:shot}. Second, the way data is delivered poses substantial challenges from an analysis perspective. Each source can use different file formats and structures. Moreover, they use different conventions about units of measure (e.g., local time vs. UTC, meters vs. yards), how the pitch is represented (e.g., coordinate system, playing direction), and how they refer to entities such as players and stadiums.  
Finally, the data sources are often collected independently of one another and it is highly non-trivial to align the data. On the one hand, it is necessary to cope with the same player being referred to using multiple naming conventions. On the other hand, time and/or location synchronizing of event and tracking data requires accounting for issues such as different clocks, different start times, and different coordinate systems~\cite{AnzerxG2021,VanRoy2023:ETSY}.  Alas, addressing all these challenges requires a substantial investment in terms of both time and money. Consequently, data is often not exploited to the fullest extent possible. 

The goal of this work is to propose a \emph{Common Data Format (CDF)} for football data that provides a standardized way to represent data arising from football matches. The CDF has been developed with analysis in mind and aims to be (1) unambiguous by proposing a common schema, (2) complete by including the necessary information for analysis tasks,  and (3) easily extendable to incorporate new data sources.  
Establishing a CDF confers many benefits as it should help alleviate some of the aforementioned pain points. First, it promotes interoperability which allows developers to build upon established standards as opposed to defining their own formats. This promises significant savings in development and maintenance costs.
Second, it reduces the barrier to entry by eliminating the technical and low-level work needed to program complex data manipulations.  Consequently, it enables less technically skilled individuals to work with data. 
As interoperability allows for more efficient use of data and computational resources, it also opens up a broader range of applications. Third, it will reduce errors introduced due to the transformations and parsing needed to align and ingest the data. This, in turn, makes it easier to spot errors in data and implement automated testing routines. 
An informed decision has been made in this first iteration to omit metrics that are not fully objective (such as but not limited to subjective event data types, possession phases or individual vendor metrics). The authors acknowledge that this may have a larger impact on some user groups (likely practitioners) more than others but see it as critical to start with an objective basis. The CDF has been designed so that users are not hindered from adding them as optional. As concepts evolve over time and become more standardised, they can be added to future editions of the CDF.

\section{Data about Football Matches}

While there are a variety of sources of data collected about sports, we will discuss six broad categories:  match sheet data, video data, event data, tracking data, match meta data and physical data. All of these data sources are of importance when thinking about a common data format.  

\subsection{Match Sheet Data}
\label{subsec:official_match_data}
Each league or federation requires collecting a minimal amount of information about each match. This information is collected for essentially all matches, whether they are professional or amateur.  This includes information about the teams (teams names, home/away, coaches' names), players (starting lineups, substitutions), referees (names), and relevant events for the competition that occur during the match (goals, cards). It also includes the final result. 

\subsection{Video Data}
\label{subsec:video_footage}
One of the oldest and most important types of data about football matches is raw video footage. Traditionally, there were two main types of video: 
\begin{description}
    \item[TV-footage] Broadcasters employ a combination of multiple cameras to capture a variety of different shots with the aim of producing a compelling TV program. The cameras are positioned to track game play while allowing for zoom-ins and wide-angle perspectives. Moreover, the cameras are also positioned such that they can capture the stadium itself as well as the team benches (e.g., to capture reactions to significant events in the match).  
    \item[Scouting or tactical feed] is a camera (or small number of cameras) that are positioned such that all 20 field players are always in view. Typically, the camera is positioned from a high perspective at the extension of the midline of the pitch. Moreover, during the  match it must be manually or automatically moved to ensure that players remain visible. 
\end{description}
Currently, top level matches are now often filmed from a variety of different perspectives:\textit{Behind goals}, \textit{coaches perspectives}, \textit{16 m line view}, and many more. 

\subsection{Event Data}
\label{subsec:event_data}
Event data or play-by-play data is a common type of data that breaks the game down into individual parts and is commonly collected in team sports such as football, ice hockey, rugby or basketball  \cite{Vracar2016}. This type of data records specific semantically meaningful events that occur during a match \cite{Pappalardo2019a} and allows users including media outlets to query particular moments of the match. For example, relevant events in football include passes, tackles, and shots. Each event is annotated with information such as the players involved, the location on the pitch where the event occurred, the time of the event in the match, and the outcome of the event.  Event data has traditionally been collected by companies such as Stats Perform\footnote{Statsperform LLC, Chicago, \url{https://www.statsperform.com/}.; former AMISCO, Prozone and Opta.}, Sportec Solutions\footnote{Sportec Solutions AG, Munich, Germany \url{https://www.sportec-solutions.de/index.html}.}, and Statsbomb\footnote{Statsbomb Services Limited \url{https://statsbomb.com/}.}, who typically use human annotators to collect this data. Each company has their own catalog of events that they collect thus meaning datasets from different vendors do not always correspond. This is particularly true for events like tackles, interceptions or clearances where observational collection methods will lead to differences. With the launch of the FIFA Football Language\footnote{FIFA, Zürich, \url{https://www.fifatrainingcentre.com/en/resources-tools/football-language/}} in 2022, a first framework has been offered to harmonise the terms and definitions used in football but more work is needed to ensure events are unambiguous. The development of automated event data detection from tracking data or video footage 
is another ongoing effort to ensure more uniform collection \cite{FIFA-Event-Detection-2022}.


The standardization of event data is important for two reasons. First, it is widely available as vendors collect data on hundreds of games per week. Second, because event data captures the semantics of the what happened in a match, they form the backbone of many different analytical tasks including match analysis, scouting or media purposes. Therefore coaches, analysts, scouts and commentators should be able to rely on a single source of truth.

\subsection{Tracking Data}
\label{subsec:tracking_data}

Tracking data record the positions of the players and the ball multiple times per second \cite{Taberner2020}. The number of recordings per second is often referred to as the frame rate. Initial tracking systems recorded the location of a player (or referee) in 2D (i.e., x/y plane) based on the player's center of mass, or center of trunk (in the case of upper mounted wearables). Henceforth this is referred to as center of body, as the player's mass is rarely if ever calculated and thus is biomechanically inappropriate to use in this context. The position of the ball is often recorded with an additional z-coordinate that provides its height relative to the pitch surface.  Newer approaches also record 3D positions for players, often augmented to include the location of various limbs (further referred to as landmarks) such as a player's head, hands, feet, hips etc. This data is sometimes referred to as skeletal tracking data, or limb tracking data. 
Tracking data can be acquired in several different ways:
\begin{description}
    \item[Fixed Cameras]  Providers can install cameras in fixed locations in the stadium, typically in its roof. This is done in such a way to ensure a high coverage (i.e., that all players are almost always in view of one camera). Such data is typically not shared outside of a specific competition such as a particular league or tournament (Champions League, World Cup). 
    \item[Mobile Cameras] For cup-competitions, national team matches, or stadiums without an appropriate roof infrastructure, mobile tracking systems are installed for single matches. Such camera systems are typically installed in central positions (extension of the midline) at the highest possible spot on the tribune (e.g., in press-boxes, on scaffolding).
    \item[Broadcast Tracking] Positions are collected by analyzing data collected for broadcast. This has the advantage of having a wider availability, i.e., it is possible to acquire for most matches without requiring exclusive access to stadiums. But it has the disadvantage that not all players are visible (i.e., in frame) all the time. Moreover, some moments of the match may lack tracking data due to the broadcaster showing a replay, stadium shot, or close up of a player. These disadvantages are sometimes overcome by exchanging broadcast footage for in stadium wide-angle scouting feed (although this can introduce other issues, such as more difficult jersey number recognition), or by bespoke software that predicts the location of players that are not in frame. 
    \item[Local/Global Positioning System (LPS/GNSS)] GNSS-wearables track players by using satellite signals to determine their position relative to the earth surface (i.e., longitude and latitude coordinates). LPS uses ground-based receivers and radio frequency signals emitted by player-worn tags for more precise positions relative to these anchors (i.e. relative to the pitch surroundings). GNSS is versatile and widely used especially for physical data, while LPS offers higher accuracy and is better suited for detailed analysis in controlled environments. Within the last FIFA World-Cups, LPS systems has especially established accurate ball tracking.    A benefit of both systems is that players are clearly identified all the time. However, especially GNSS systems require a direct visibility to a certain number of satellites, which cannot be ensured across all football stadiums.

    %
 
\end{description}
Sometimes a combination of the two systems  has been used. For example FIFA deployed LPS for the ball and fixed-cameras for players for the 2022 Men's and the 2023 Women's World Cup. 

Multiple studies have investigated the validity of specific tracking data systems (e.g., \cite{Linke2020}). However, these tests cannot be conducted during official matches. Hence, it is difficult to fully quantify their accuracy.

\subsection{Match Meta Data}
\label{subsec:meta_data}
Match meta data contain relevant information that describe the context of the match. Such information includes basic player and team information, kick off time, the type of match (e.g., league, domestic cup, friendly), and venue information (pitch size, turf, stadium identifier, attendance). Beyond this basic information, it can include relevant external factors such as the weather.  




\subsection{Player Physical Data}
Players also wear a number sensors during training sessions and matches to track their physical status.
Most notably, the often wear a GNSS device that also contains an inertial measurement unit (e.g., accelerometer and gyroscope). Furthermore, it is possible that athletes wear other devices such as a heart rate monitor. These systems are used to derive a number of parameters such as the total distance covered and the number of accelerations. These are also typically subdivided by defining different intensity zones.  A number of companies make these devices including Catapult Sports\footnote{Catapult Sports, Melbourne, \url{https://www.catapult.com/}} and STATSports\footnote{STATSports Group Limited, Ireland, \url{https://pro.statsports.com/}}. 
Moreover, athletes are often asked to fill in questionnaires~\cite{Buchheit:2016} that might include information about their sleep or the rating of perceived exertion (RPE)~\cite{Borg1982} for a training session.

A number of the recorded parameters (e.g., distance covered, accelerations) can also be derived from tracking data. 
 However, most teams do not have an optical or LPS tracking system installed on training pitches which leads to the use of the body-worn sensors as well. 
 


\section{Challenges Posed by Football Data}

Football data presents a myriad of challenges for data analysts, data scientists, and data engineers. These challenges can be broadly classified into four categories. The first category pertains to the differences in data specifications among data providers. What kind of data does each provider collect? The second category relates to the differences in data representation among data providers. How does each provider represent their data? The third category involves the differences in data delivery methods among data providers. How does each provider deliver their data? The last category describes the data quality. What kind of errors can result from manual data collection processes? 

\subsection{Data Specification}

Data providers employ varied data specifications when collecting event data. 
Consequently, the event data collected by one provider for a specific game can markedly differ from the data collected by another provider for the identical game. 
An example could be a definition of a cross. One vendor requires a cross to come from a specific area on the pitch with a non-zero $z$ height for the ball, while another vendor can describe a cross as a pass simply being played forward with some arbitrary distance and angle into the opponents box. 

The discrepancies in event data across different providers are typically attributed to variations in how the game is broken down into individual events and the definitions applied to those events. Furthermore, the scope of event data also diverges among providers, with some incorporating more derived data into their data feeds than others.

Data providers independently determine the set of events they wish to collect. The International Football Association Board (The IFAB)\footnote{IFAB, Zürich, \url{https://www.theifab.com/organisation/}}, serving as the independent authority of the Laws of the Game, only outlines events that are vital for upholding the game’s rules. These events include instances such as kick-offs, free kicks, goal kicks, and offside rulings. However, this limited set of events is insufficient for a comprehensive analysis of a football game. Therefore, data providers must independently establish an appropriate set of events for their collection purposes.

The set of events collected can greatly differ from one data provider to another. These differences are primarily due to three factors. The first factor is the tactical trends in the region where the data provider operates. For example, a data provider based in Europe might focus more on collecting events related to players pressuring the ball carrier, as pressing-based playing styles have gained traction in Europe. The second factor is the intended audience for the data feed. Data providers who cater to media companies and betting agencies have different priorities than those serving clubs and associations. The third factor is the complexity and the necessary time commitment required to collect certain types of events~\cite{bialik:annotate}. 
Data providers gather varying sets of events, but most tend to collect the most basic event types like shots and passes, which constitute the majority of on-ball events. However, since events like passes are not specified by the Laws of the Game\footnote{IFAB, Laws of the Game, \url{https://digitalhub.fifa.com/m/1cf301829f1cf996/original/ifab-laws-of-the-game-2020-21.pdf}}, data providers resort to their own definitions. Consequently, specific actions on the pitch may be classified differently by different providers, leading to inconsistencies in the resulting statistics and metrics~\cite{bialik:shot}.

Furthermore, the breadth of the data specification also varies among data providers. Some providers limit their collection to actions performed by players, while others extend their scope to include meta events, such as changes in temperature or pitch conditions. Additionally, certain data providers incorporate performance metrics into their data specification, such as expected-goals values for shots  \cite{AnzerxG2021}, expected pass values for passes \cite{Anzer2022_xPass} or metrics on pressure \cite{Andrienko2017}, while others choose not to.

\subsection{Data Representation}
\label{sec:datarep-problems}

In addition to the specific data that each provider collects, the representation of the data also varies among providers. Specifically, data providers

\begin{itemize}
  \item utilize \textbf{different coordinate systems}. For example, one provider may place the origin of the coordinate system at the center of the pitch, while another may place it at a corner of the pitch.
  \item  adopt \textbf{different pitch coordinates}. Some providers use normalized pitch coordinates, while others use actual pitch coordinates. The normalization of pitch coordinates introduces further challenges \cite{Brandes-2023-Pitch}, as pitches can vary in size, while most pitch markers are of fixed dimensions. For example, the penalty area and the penalty spot’s distance from the goal remain constant, regardless of the pitch dimensions.
  \item  apply \textbf{different pitch orientations}. For instance, some providers use the actual playing direction for each team, while others adjust pitch coordinates so that each team appears to be attacking the same goal, typically from left to right.
  \item employ \textbf{different units of measurement}. For instance, one provider may use metric units such as meters, while another may use imperial units such as yards to denote distances.
  \item use \textbf{different date and time notations}. Some providers use Coordinated Universal Time (UTC), while others only report local time or only game time (e.g., period, minute and second of play).
  \item assign their \textbf{unique identifiers} to entities such as players, teams, matches, officials, and competitions.
\end{itemize}

\subsection{Data Delivery}

Data providers not only vary in terms of data specification and data representation, but also in their choice of data formats and data delivery methods. Specifically data providers, 

\begin{itemize}
  \item employ \textbf{different file formats}. While most providers offer the option between JSON and XML, some providers exclusively support one over the other.
  \item adopt \textbf{different file structures}. For example, some providers favor flat file structures, while others opt for deeply-nested file structures.
  \item distribute their data through \textbf{various delivery methods}. Although most providers distribute their data via a traditional REST API, some providers also propose alternative delivery methods such as a message broker (e.g., RabbitMQ in the case of Hawk-Eye) or an FTP server.
\end{itemize}

Additionally, some data providers offer their \textbf{own coding libraries} (in the form of Python or R packages) to more quickly get started ingesting their data. 
\subsection{Data Quality}

The mostly manual collection process of event data can lead to various errors that are later delivered to the customers. Providers have varying degrees of safety measures built into their process to reduce these, but even with them in place only a fraction can be captured. While virtually every part of the manual collection part could at some point deliver erroneous data, the following are the most frequent errors:

\begin{itemize}
\item Some \textbf{attributes of events are incorrectly captured or incorrectly ordered}: These can include \textbf{timestamps}, \textbf{coordinates} or even the involved \textbf{player(s)}. As collecting all events during a game is a fast paced challenge it is nearly impossible to always be recognizing the right player(s), clicking on the exact location on the pitch or stopping the video at the exact right time. 
\item \textbf{Meta information may be false or missing}: when inserting or altering the meta data base errors can occur regarding birth dates, names or other additional information.
\item Entire \textbf{events can be missing}: this can be either an oversight by the human operator, or for the most part, this is a result of using broadcast video (and not in venue or of a 'clean' video feed) during the collection process. These broadcast videos typically include replays, zoom-ins or other reasons, why the occurring events during the game are not being shown in their video. Therefore, it is nearly impossible for the collectors to estimate what may have happened in the mean time, while the relevant part of the pitch was not on display.
\end{itemize}

Apart from these obvious errors there is also some uncertainty in events. Since most of the definitions are not absolute, but rather involve some form of human judgment. Even within the same provider, a very similar situation may classified differently for no apparent reason.


\section{Related Work}
Setting data standards and introducing a standardized data format is not a unique idea in the football landscape. The necessity of both is widely acknowledged, as it is also a cornerstone of the Federal Data Strategy 2020 Action Plan of the U.S. government.\footnote{\url{https://resources.data.gov/standards/concepts/}} Similar initiatives are currently establishing such formats in fields such as medicine, healthcare, water management, and transportation. 
One success story is the General Transit Feed Specification (GTFS),\footnote{\url{https://gtfs.org}} an open standard used to distribute relevant information about transit systems to commuters. By many transport agencies adhering to this standard, commercial use by different companies has been enabled, incorporating information into products like Google, Apple, Transit App, Open Trip Planner, and even apps created by riders. 
This example highlights some of the benefits of a common data format.

Although football data currently lacks a consensus and standard, there have been several individual attempts to simplify analysis by proposing unifying data representations. Interestingly, these are often backed by open-source software packages with permissive licenses (i.e., commercial use is allowed). The SPADL (Soccer Player Action Description Language)~\cite{Decroos2019} proposes a tabular representation for a subset of the key information recorded in event data. This puts the data in a natural format and facilitates computing advanced performance metrics from event data such as expected goals (xG)~\cite{robberechts2020:xg} and Expected Threat (xT)~\cite{rudd2011framework,Singh:xT}. The software package supports parsing data from several different data providers and this package is used in both research and industry. 

For tracking data, FIFA and F.C. Barcelona jointly created global standards for Electronic Performance and Tracking Systems (EPTS)~\cite{fifa:epts}.  More recently, there are open-source packages such as kloppy~\cite{kloppy} and floodlight \cite{raabe2022floodlight} that support parsing both tracking and event data. Moreover, there are several open-source algorithms for synchronizing tracking and event data~\cite{oonk:databallpy,VanRoy2023:ETSY}.

 While these initiatives are acknowledged by data providers and widely used by researchers and practitioners, an effort to harmonize all match data, supported by a large group of stakeholders has not yet culminated. 
  Similar to the work of Robertson et al.~\cite{robertson2023development}, the CDF aims to provide a framework for handling football data that  offers possible solutions to common challenges as outlined in the previous section.

\section{The Common Football Data Format}
The goal of the CDF is to provide a uniform and standardized format for data generated from football matches. The guiding principles of the CDF are to promote the ease of analysis of football data by being:
\begin{description}
    \item[Unambiguous] The format strives for uniformity and clarity by providing a common schema. Moreover, it ensures that sufficient \emph{context} is available to understand the provenance of the data. 
    \item[Sufficient] The focus is on ensuring that enough information is provided to enable common downstream analysis tasks in areas such as scouting and match preparation. 
    \item[Extendable] The football data landscape is not static, and it is important for the format to be able to accommodate new innovations.
\end{description}
In essence, the CDF aims to present a precise and well-defined interface to football data that mitigates the existing pain-points associated with the current data landscape. Specifically, the CDF targets clubs, federations and researchers. It aims to allow them  to easily build their own logic and applications on top of provided data without having to contend with some of the (arbitrary) design choices made by data collectors.  
Having vendors supply data in the CDF will help shorten the time between data exploration and delivery of actionable products to stakeholders as well as facilitate more impactful collaborations with researchers and developers.


This section presents {\bf Version 1.0.0} of the CDF. It describes the key properties of the CDF and discusses some of the design choices that were made. First, we will outline the technical specifications of the CDF in terms of what data sources are included and within each data source what information should be provided. Second, we will discuss some of the representational choices that were made to help ensure the clarity of the provided data.  Third, we will propose a concrete way to deliver data in the CDF to enable easy ingestion and analysis of the data. Finally, we would like to highlight that this is the first version of the CDF and it will continue to evolve as the data landscape changes.

\subsection{Specification of the CDF Schema}
The CDF specifies a minimal schema for five types of match data: match sheet data, video footage, event data, tracking data, and match meta data. However, it is not mandatory that all five types of data are available. The only two requirements on availability are that (1) match sheet data must be included, and (2) if either event or tracking data is provided, then match meta data must also be included. Additionally, if both event and tracking data are provided, these data sources should be synchronized in time and location (see e.g., \cite{oonk:databallpy,VanRoy2023:ETSY,AnzerxG2021,Anzer2022_xPass}).

We now describe what fields must be collected for each data source, which are shown in bold in the corresponding tables. 
\begin{description}
    \item[Match sheet data] As shown in Table~\ref{tab:cdf-matchdata-required}, the CDF collects the standard match information that is needed to manage a competition (e.g., league) and report a box score.
    \item[Video footage] Required meta information for the video footage are listed in Table~\ref{tab:cdf-video}. Information like period starting whistles simplify video automation processes in combination with event and tracking data.
    \item[Event data] Table~\ref{tab:cdf-event-required} provides the schema for event data. The CDF only includes events that are either objective or whereby clear consensus exists on their definition (irrespective of the collection method): Events that do not meet this threshold are excluded from the CDF.  The events that meet this requirement are listed in event\_(sub)type fields of the provided schema.\footnote{There is ongoing work from a FIFA expert group that is developing a touched-based event model~\cite{FIFA_2023_Whitepaper}. } 
    
    \item[Tracking data] Table~\ref{tab:cdf-com-tracking-required} shows the information that is required for basic center of body tracking.  Beyond the player coordinates, each frame of the tracking data should be augmented with information about which team has the ball and whether the ball is in or out of play.  Every period can take on the values first half, second half, first half of overtime, second half of overtime, and penalty shootouts. Frame identifiers  should be monotonically increasing unique integers starting at 0. If the ball is tracked at a different sampling rate, then the provider should synchronize this to the player sampling rate and only provide these measurements in the schema outlined in Table~\ref{tab:cdf-com-tracking-required}. In this situation, the provider can also provide an additional, separate file that omits player positions and just reports the ball location at its (higher) sampling rate. If skeletal tracking is provided, a separate schema shown in Table~\ref{tab:cdf-limb-tracking-required} is used for the joint locations. A separate schema is used for ease of processing.
    \item[Match meta data] Table~\ref{tab:cdf-meta-required} lists what will be included in the meta data. Beyond contextual information about the match (e.g., teams, players, location, data, and start time), this includes information that is often needed for subsequent analysis. For example, it records detailed information about what data (e.g., tracking) were collected during the match and how the data were collected (e.g., vendor, version of the software, frame rate). Such information is crucial for understanding the provenance of the data. Finally, there is field to denote which version of the CDF the data conforms to. 
\end{description}
Beyond these mandatory fields, Appendix~\ref{app:optional} discusses a list of optional fields for each data source.  If such information is collected, then it must conform to the provided field names and possible values the field can take on.  Note that optional fields are extendable: vendors may provide any additional information that deem to be of relevance and interest. However, this should not circumvent the required or optional fields. 

\subsubsection*{Versioning}
As the CDF will evolve over time, we will maintain semantic version numbers of the form Major.Minor.Patch. In this system, Major, Minor and Patch are integers. Within a fixed Major, we will ensure backwards compatibility. 

\subsubsection*{Schema Validation Tool}
To aid developers with the implementation of the schema's described in the subsequent sections, the CDF is accompanied by a \href{https://pypi.org/project/common-data-format-validator/0.0.3/}{Common Data Format Validator Python package} that can be used to validate their JSON schemas. This Python package additionally contains a changelog where all changes made to the CDF are recorded.

\begin{table}[H]
    \caption{Description of the mandatory Match Sheet Data CDF fields. Any description ending with a $^\star$ is repeated once per player. Only players named to the match day squad are listed. Any description ending with a $^\dagger$ is repeated once per event. Note that \{i\} is used as an index for the items in a list. Figure~\ref{fig:match-data-json} describes the specific format for providing results.}
    \label{tab:cdf-matchdata-required}
    \begin{adjustbox}{max width=\textwidth}
    \begin{tabular}{lp{2.75cm}p{7cm}l}
    \hline
    Root & Field & Description & Type \\ \hline
    match & id & Unique match identifier & String \\
    match/status & is\_neutral & Denotes whether the game was hosted in a neutral venue (true) or not (false) & Boolean \\ 
    match/status & has\_extratime & Denotes whether the game went to extra time (true) or not (false) & Boolean \\
    match/status & has\_shootout & Denotes whether the game had a penalty shootout (true) or not (false) & Boolean \\
    match/result/final & \{home\text{\textbar}away\} & Result after the final whistle excluding penalty shootout goals (i.e. home goals, away goals) & Integer \\
    match/result/final & winning\_team\_id & Unique identifier of the winning team & String \\
    match/result/first\_half & \{home\text{\textbar}away\} & Result after the first half (i.e. home goals, away goals) & Integer \\
    match/result/second\_half & \{home\text{\textbar}away\} & Result after the second half (i.e. home goals, away goals) & Integer \\
    match/result/first\_half\_extratime & \{home\text{\textbar}away\} & Result after the first half of extra time (i.e. home goals, away goals) & Integer \\
    match/result/second\_half\_extratime & \{home\text{\textbar}away\} & Result after the second half of extra time (i.e. home goals, away goals) & Integer \\
    match/result/shootout & \{home\text{\textbar}away\} & Score for the penalty shootout (i.e. home goals, away goals, shootout goals only) & Integer \\
    teams/\{home\text{\textbar}away\} & id & Unique identifier for the home or away team & String \\
    teams/\{home\text{\textbar}away\}/players/\{i\} & id & Unique player identifier$^\star$ & String \\
    teams/\{home\text{\textbar}away\}/players/\{i\} & first\_name & First name$^\star$ & String \\
    teams/\{home\text{\textbar}away\}/players/\{i\} & last\_name & Last name$^\star$ & String \\
    teams/\{home\text{\textbar}away\}/players/\{i\} & team\_id & Unique team identifier denoting the team the player plays for$^\star$ & String \\
    teams/\{home\text{\textbar}away\}/players/\{i\} & jersey\_number & Jersey number for a player$^\star$ & Integer \\
    teams/\{home\text{\textbar}away\}/players/\{i\} & is\_starter & Denotes whether a player started the game (true) or not (false)$^\star$ & Boolean \\
  teams/\{home\text{\textbar}away\}/players/\{i\} & has\_played & Denotes whether a player played in game (true) or not (false)$^\star$ & Boolean \\
    referees/\{i\} & id & Unique referee identifier & String \\
    events/goals/\{i\} & time & Time a player scored$^\dagger$ & String \\
    events/goals/\{i\} & player\_id & Identifier of the player who scored$^\dagger$ & String \\
    events/goals/\{i\} & assist\_id & Identifier of the player who assisted$^\dagger$ & String \\
    events/goals/\{i\} & team\_id & Identifier of the team that scored (for own goals this should be identifier of the team that gained a goal). $^\dagger$ & String \\
    events/goals/\{i\} & is\_own\_goal & Denotes whether it was an own goal (true) or not (false)$^\dagger$ & Boolean \\
    events/goals/\{i\} & is\_penalty & Denotes whether it was a penalty (true) or not (false)$^\dagger$ & Boolean \\
    events/goals/\{i\}/score & \{home\text{\textbar}away\} & Team score after the goal$^\dagger$ & Integer \\
    events/substitutions/\{i\} & in\_time & Time in UTC a player is substituted in$^\dagger$ & String \\
    events/substitutions/\{i\} & in\_player\_id & Identifier of the player that is substituted out$^\dagger$ & String \\
    events/substitutions/\{i\} & out\_time & Time in UTC a player is substituted in$^\dagger$ & String \\
    events/substitutions/\{i\} & out\_player\_id & Identifier of the player that is substituted out$^\dagger$ & String \\
    events/substitutions/\{i\} & team\_id & Identifier of the team that made the substitution$^\dagger$ & String \\
    events/cards/\{i\} & time & Time in UTC a player received a card$^\dagger$ & String \\
    events/cards/\{i\} & player\_id & Identifier of the player who received a card$^\dagger$ & String \\
    events/cards/\{i\} & type & Type of card which can be yellow\_card, red\_card or second\_yellow\_card$^\dagger$ & String \\
    events/cards/\{i\} & team\_id & Identifier of the team that made the received a card$^\dagger$ & String \\
    meta & vendor & Match sheet data vendor name (e.g. "company\_a") & String \\    
    \end{tabular}
    \end{adjustbox}
\end{table}

\begin{table}[H]
\centering
\caption{Description of the Video Footage CDF fields.}
\label{tab:cdf-video}
\begin{adjustbox}{max width=\textwidth}
\begin{tabular}{lp{2.5cm}p{8cm}l}
\hline
Root & Field & Description & Type \\ \hline
match & id & Unique match identifier & String \\
recording & fps & Frames per second (i.e. frame rate) from vendor & Integer \\
recording & resolution & Resolution of the video in pixels (e.g., 3840x2160, 1920x1080, ...) & String \\
recording & start\_time & The start time in UTC of the recording & String \\ 
recording & operation\_type & Information whether the camera operation was manual or automated & String \\
recording & recording\_type & Information how the video was recorded (e.g. fixed camera, camcorder,...) & String \\
recording & perspective & Camera angle (tactical\_wide, camera\_1, high\_behind\_right, high\_behind\_left, cable\_camera, 16m\_right, 16m\_left, broadcast) & String \\
whistles/\{i\} & type & Whistles that start and end major periods of play such as the start and end of halves and interruptions (e.g., weather, VAR review, player health events, streakers or abandoned). Examples of types first\_half, second\_half, weather\_delay, health\_delay, injury\_treatment fan\_health\_delay etc. & String \\ 
whistles/\{i\} & sub\_type & Sub type related to an interruption, for example start or end. & String \\
whistles/\{i\} & time & The time in UTC of the whistle  & String \\
whistles/\{i\} & video\_time & The time tag of the whistle in milliseconds  & Integer \\ \hline

\end{tabular}
\end{adjustbox}
\end{table}

\begin{table}[H]
\centering
\caption{Description of the mandatory Event CDF fields. Every recorded event must include all these fields. The coordinate system is defined in Paragraph~\ref{p:coords} where the x axis is the sidelines and the y axis the goal lines. Note that \{i\} is used as an index for the items in a list.}
\label{tab:cdf-event-required}
\begin{adjustbox}{max width=\textwidth}
\begin{tabular}{lp{2.5cm}p{8cm}l}
\hline
Root & Field & Description & Type \\ \hline
match & id & Unique match identifier & String \\
meta & is\_synced & Indicates if synced tracking data is available (true) or not (false) for this event & Boolean \\
event & id & Unique identifier of the event & String \\
event & time & Absolute time in UTC of when the event started (e.g., moment the pass is given, the moment the ball leaves the hands on a thrown in) & String \\
event & period & Period of the match which can be first\_half, second\_half, first\_half\_extra, second\_half\_extra, or shootout & String \\
event & type & Name of the event type (e.g. shot, pass, referee, misc, etc.) & String \\
event & sub\_type & Name of the event subtype, which can be for shot - (None, penalty\_kick, free\_kick, corner\_kick); pass - (None, throw\_in, free\_kick, corner\_kick, goal\_kick, kick\_off); referee - (final\_whistle, foul, caution, offside, substitution, player\_on, player\_off. Player on and player off events occur for example when a player has to abandon the game due to a lack of available substitutions, or as a temporary measure after a medical check); misc - (other\_ball\_action, chance\_without\_shot, tackle) & String \\
event & is\_successful & Denotes whether the event was successful (true) or not (false) & Boolean \\
event & outcome\_type & Detailed event outcome options: shot - (successful, saved, blocked, wide, woodwork, own\_goal); pass - (e.g. successful, out\_of\_play, intercepted);  referee - (start, end, injury, yellow\_card, red\_card or second\_yellow\_card); misc - (e.g. successful, unsuccessful) & String \\
event & player\_id & Unique identifier of the player performing the action. For example, player committing a pass, or making a tackle. & String \\
event & team\_id & Unique team identifier of the player performing the action & String \\
event & receiver\_id & Unique identifier of the player receiving a pass. Leave \textit{null} when the event is not a pass or when the pass has no receiver. & String \\
event & receiver\_time & Absolute time in UTC of the first moment the ball was received. Leave \textit{null} when the event is not a pass or when the pass has no receiver. & String \\
event & x & x location where the action of player\_id happened (m).   & Float \\
event & y & y location where the action of player\_id happened (m).& Float \\
event & x\_end & x location where the action of player\_id ended (m).   & Float \\
event & y\_end & y location where the action of player\_id ended (m). & Float \\
event & body\_part & Denotes the body part used by player\_id which can take on the values left\_foot, right\_foot, head, or other & String \\
event & related\_event\_ids & Unique identifier(s) of the events related to the action, or pass and associated receival event. For example, a related aerial duel event, the receival event related to a pass, or the player\_off event associated with a player\_on event. Leave null if no related events exist. & Array[String] \\
\end{tabular}
\end{adjustbox}
\end{table}

\begin{table}[H]
\centering
\caption{Description of mandatory center-of-body tracking CDF fields recorded per frame. Any description ending with a $^\star$ is repeated once per player per frame. The coordinate system is defined in Paragraph~\ref{p:coords} where the x axis is the sidelines and the y axis the goal lines. Note that \{i\} is used as an index for the items in a list.}
\label{tab:cdf-com-tracking-required}
\begin{adjustbox}{max width=\textwidth}
\begin{tabular}{lp{1.5cm}p{8cm}l}
\hline
Root & Field & Description & Type \\ \hline
& frame\_id & Unique frame identifier & Integer \\
& timestamp & Timestamp of the frame in UTC  & String \\
& period & Period of the match (first\_half, second\_half, first\_half\_extratime, second\_half\_extratime, shootout) & String \\
match & id & Unique match identifier & String \\
teams/\{home\text{\textbar}away\} & id & Unique identifier for the home or away team$^\star$ & String \\
teams/\{home\text{\textbar}away\}/players/\{i\} & id & Unique identifier for a player$^\star$ & String \\
teams/\{home\text{\textbar}away\}/players/\{i\} & x & x location of the player on the pitch (m)$^\star$. & Float \\
teams/\{home\text{\textbar}away\}/players/\{i\} & y & y location of the player on the pitch (m)$^\star$.  & Float \\
ball & x & x location of the ball on the pitch (m). & Float \\
ball & y & y location of the ball on the pitch (m).  & Float \\
ball & z & z location of the ball on the pitch (m).  & Float \\
\end{tabular}
\end{adjustbox}
\end{table}

\begin{table}[H]
\centering
\caption{Description of mandatory Skeletal Tracking CDF fields recorded for each frame. Any description ending with a $^\star$ is repeated once per player per frame. Any description ending with a $^\diamond$ is repeated once per player per frame per landmark (e.g knee\_left, hip\_right, ear\_right etc.). Note that \{i\} and \{j\} are used as indices for items in lists. Landmark names should follow snake-case with any direction indicator as a suffix. }
\label{tab:cdf-limb-tracking-required}
\begin{adjustbox}{max width=\textwidth}
\begin{tabular}{lp{1.5cm}p{8cm}l}
\hline
Root & Field & Description & Type \\ \hline
& frame\_id & Unique frame identifier & Integer \\
& timestamp & Timestamp of the frame in UTC  & String \\
& period & Period of the match (first\_half, second\_half, first\_half\_extra, second\_half\_extra, shootout) & String \\
match & id & Unique match identifier & String \\
teams/\{home\text{\textbar}away\} & id & Unique identifier for the home or away team$^\star$ & String \\
teams/\{home\text{\textbar}away\}/players/\{i\} & id & Unique identifier for a player$^\star$ & String \\
teams/\{home\text{\textbar}away\}/players/\{i\}/landmarks/\{j\} & index & Unique identifier for a landmark$^\diamond$ & String \\
teams/\{home\text{\textbar}away\}/players/\{i\}/landmarks/\{j\} & name & Name for a landmark$^\diamond$ & String \\

teams/\{home\text{\textbar}away\}/players/\{i\}/landmarks/\{j\} & x & Relative x coordinate of landmark in relation to the point of origin (m)$^\diamond$  & Float \\

teams/\{home\text{\textbar}away\}/players/\{i\}/landmarks/\{j\} & y & Relative y coordinate of landmark in relation to the point of origin (m)$^\diamond$  & Float \\
teams/\{home\text{\textbar}away\}/players/\{i\}/landmarks/\{j\} & z & Relative z coordinate of landmark in relation to the point of origin (m)$^\diamond$  & Float \\
teams/\{home\text{\textbar}away\}/players/\{i\}/landmarks/\{j\} & children & List of children indexes associated with keypoint$^\diamond$ & Array[Int] \\
teams/\{home\text{\textbar}away\}/players/\{i\}/landmarks/\{j\} & is\_visible & If landmark is detected (true) or inferred (false) $^\diamond$ & Boolean \\

\end{tabular}
\end{adjustbox}
\end{table}

\begin{table}[H] 
\caption{Description of mandatory Match Meta CDF. Any description ending with a $^\star$ is repeated once per player. Note that \{i\} is used as an index for the items in a list. Separate ball version, name, fps and collection timing is only required when providing an independent ball file (i.e. when the ball sampling rate is different).}
\label{tab:cdf-meta-required}
\begin{adjustbox}{max width=\textwidth}
\begin{tabular}{lp{2.5cm}p{8cm}l}
\hline
Root & Field & Description & Type \\ \hline
competition & id & Unique identifier for the competition & String \\
season & id & Unique identifier for the season & String \\
match & id & Unique identifier for the match & String \\
match & kickoff\_time & Scheduled kickoff time in UTC & String \\
match/periods/\{i\} & type & Possible options are first\_half, second\_half, first\_half\_extratime, second\_half\_extratime, shootout & String \\
match/periods/\{i\} & play\_direction & The direction of play for the home team. Possible options are left\_right or right\_left. & String \\
match/whistles/\{i\} & type & Whistles that start and end major periods of play such as the start and end of halves and interruptions (e.g., weather, VAR review, player health events, streakers or abandoned). Examples of types first\_half, second\_half, weather\_delay, health\_delay, injury\_treatment fan\_health\_delay etc. & String \\
match/whistles/\{i\} & sub\_type & Sub type related to an interruption, for example start or end. & String \\
match/whistles/\{i\} & time & The time in UTC of the whistle  & String \\
teams/\{home\text{\textbar}away\} & id & Unique identifier for the home or away team & String \\
teams/\{home\text{\textbar}away\}/players/\{i\} & & Array of player objects. One array under teams/home, one array under teams/away & Array \\
teams/\{home\text{\textbar}away\}/players/\{i\} & id & Unique player identifier$^\star$ & String \\
teams/\{home\text{\textbar}away\}/players/\{i\} & team\_id & Unique team identifier denoting the team the player plays for$^\star$ & String \\
teams/\{home\text{\textbar}away\}/players/\{i\} & jersey\_number & Jersey number for a player$^\star$ & Integer \\
teams/\{home\text{\textbar}away\}/players/\{i\} & is\_starter & Denotes whether a player started the game (true) or not (false)$^\star$ & Boolean \\
stadium & id & Unique identifier for the stadium & String \\
stadium & pitch\_length & Length of the pitch (m), unknown if not available, null if not available & Float \\
stadium & pitch\_width & Width of the pitch (m), unknown if not available, null if not available & Float \\
meta/video & perspective & Camera perspective (e.g. in\_stadium, broadcast, tactical, tactical\_wide). & String \\
meta/landmarks & nodes & Skeletal hierarchy used (i.e., what landmarks are tracked). A explicit example of how to provide this data can be found in Figure~\ref{fig:skeletal-hierarchy} & JSON Object \\
meta/\{event\text{\textbar}tracking\text{\textbar}video\text{\textbar}landmarks\text{\textbar}ball\text{\textbar}meta\text{\textbar}cdf\}  & version & Version number for the  data collection in use (e.g. "0.1.0") & String \\
meta/\{event\text{\textbar}tracking\text{\textbar}video\text{\textbar}landmarks\text{\textbar}ball\text{\textbar}meta\} & name & Vendor name of the tracking, limb tracking, event or video data & String \\
meta/\{tracking\text{\textbar}video\text{\textbar}landmarks\text{\textbar}ball\} & fps & Frames per second (i.e., frame rate) of tracking, limb tracking or video & Integer \\
meta/\{event\text{\textbar}tracking\text{\textbar}landmarks\text{\textbar}ball\} & collection\_timing & Indicates if the data was collected live or post\_match. & String \\
\end{tabular}
\end{adjustbox}
\end{table}

\newpage

\subsection{Representational Conventions in the CDF} 
\label{cdf:represent_conventions}
As highlighted in Subsection~\ref{sec:datarep-problems}, a limiting factor is the wide array of different representational conventions currently being used. To promote clarity and simplify subsequent analysis, the CDF adopts the following conventions:

\paragraph{Units of Measure} The CDF adopts metric units for all quantities. Times are recorded in Coordinated Universal Time (UTC).  

\paragraph{Pitch Coordinates}
\label{p:coords}
Because the CDF aims to be universally applicable (i.e., used in both international and domestic competitions), it cannot assume a standard pitch size of $105 \times 68$ meters.  The CDF uses absolute coordinates where 
\begin{enumerate}
    \item $(x,y,z)=(0, 0, 0)$ is the center of pitch,
    \item the $X$-axis represents the sideline. Therefore, the pitch boundaries range from 
    \begin{equation}
    \label{x:pitch}
\left[ -\frac{L_{\text{pitch}}}{2}, \frac{L_{\text{pitch}}}{2} \right], 
\end{equation}
    \item the $Y$-axis represents the endline (i.e., where the goals are placed). Therefore, the pitch boundaries range from 
    \begin{equation}
    \label{y:pitch}
\left[ -\frac{W_{\text{pitch}}}{2}, \frac{W_{\text{pitch}}}{2} \right], 
\end{equation}
\end{enumerate}

\begin{figure}[H]
  \centering
  \includegraphics[width=0.85\textwidth]{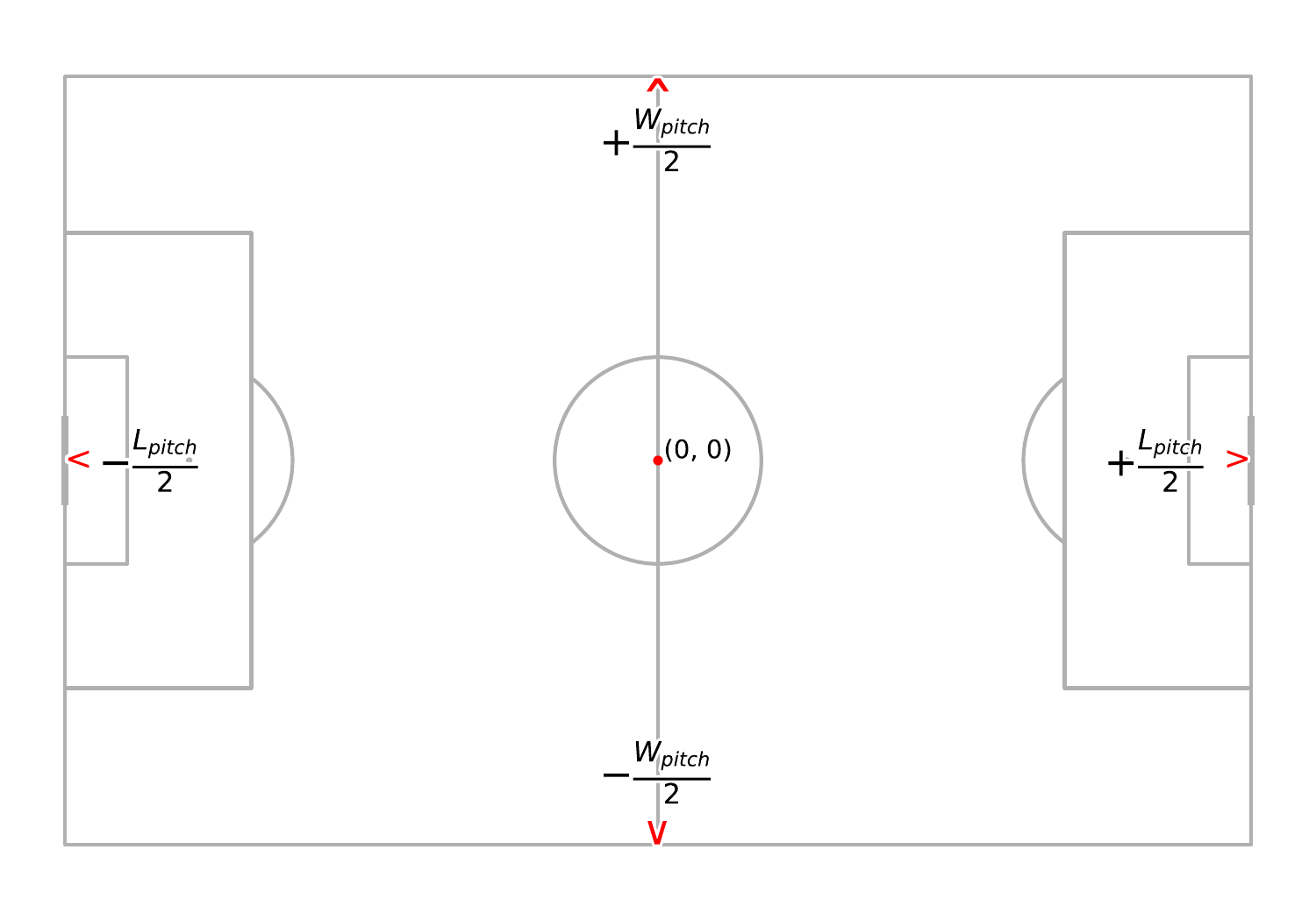}
  \caption{Pitch coordinate system}
  \label{fig:pitch-dimens}
\end{figure}

\noindent A visual representation of the pitch coordinate system described above is available in Figure~\ref{fig:pitch-dimens}.
\newline
\noindent
Note that the pitch's length ($L_{\text{pitch}}$) and width ($W_{\text{pitch}}$) are provided in the meta data. Events that take place outside of the field of play will be denoted by coordinate values that fall outside the ranges given in Equation~\ref{x:pitch} and Equation~\ref{y:pitch}.

An advantage of a coordinate system with both positive and negative values is that it allows for custom normalization simply by multiplying various coordinates by $-1$. 

\paragraph{Playing direction}
\label{sec:playing-direction}
CDF assumes that the home team \emph{always} plays from left to right for the entire match. The meta data contains sufficient information (e.g., home/away teams, direction of play in each period) to enable a user to derive alternative representations. 

In the case of penalty shootouts, the penalty taker always shoots left to right (i.e., towards the goal on the right).

\paragraph{Time}
When a time is required, the CDF works based on timestamps because they offer both clarity and flexibility for downstream applications. Moreover, timestamps gracefully enabling coping with interruptions. For example, if a match stops at 2024-08-29T14:00:00~UTC for exactly two hours for any reason, the first timestamp (e.g., in the tracking data) on its restart should then be 2024-08-29T16:00:00~UTC.

\paragraph{Unique Identifiers}

The CDF relies on a clearly defined ID-space where unique IDs are available for all entities tracked over time included matches, players, teams, venues/stadiums, competitions, coaches, and referees. The CDF simply requires that vendors have unique IDs for each of these entities.\footnote{Ideally, the CDF would require vendor-agnostic unique IDs for all entities, but such an undertaking is beyond the scope of this project. Note that FIFA has an ongoing initiative for players with FIFA Connect-IDs. } 



\paragraph{Missing Values}
Because the CDF aims for verbosity and clarity, any missing value should be explicitly denoted. Hence, the end user can decide how to deal with them. This allows for faster debugging of variables that perhaps contain bugs, both on the vendor and user side. 
For this, the CDF adopts the approach of using the reserved value of {\bf null} to represent a missing value for every data type.



\paragraph{Floating Point Precision} To avoid a false sense of precision and unnecessarily large data files, floating-point values should use the minimum number of significant digits necessary. For instance, $x$ and $y$ coordinates should be represented with a maximum of three decimal places.


\subsection{Structure of Delivered Data}

The CDF covers data that are provided both post-match and real-time (i.e., live). Therefore the CDF will use a file format that supports both use cases in order to avoid needing separate processes and file structures for handling real-time data and post-match data. Concretely, data that is not streamed (e.g., match sheet data, meta data) should be delivered in the JSON format (\emph{.json}) and real-time data (e.g., event data, tracking data) should delivered using \textit{JSON Lines} (\emph{.jsonl}).\footnote{We acknowledge that other end users such as the media may have other requirements with regards to live data delivery.}  In all cases, data should always be encoded in UTF-8. 

We have selected these formats because they confer three important benefits. First, they are relatively compact in terms of memory used per file (compared to other commonly used file formats like JSON or XML), and are also efficient in terms of maximum memory capacity used when loading or streaming into memory. Second, a well-structured (and versioned) JSON (Lines) file format will allow providers to enrich their data with more features as time progresses and new metrics become available without breaking any existing infrastructure. Third, all popular data processing tools ($pandas$, $PySpark$, $polars$ and $R$) support this file type, and therefore further post-processing can be left to each individual receiver.



\subsubsection{Meta- and Match Sheet Data File Format}
The match sheet and meta data should be delivered in a JSON file ($.json$). Figure~\ref{fig:match-data-json}  shows a simplified example of match sheet data where some mandatory CDF fields are omitted for readability. The optional fields are shown in Table~\ref{tab:cdf-matchdata-optional}. Figure~\ref{fig:meta-data} depicts a simplified example of meta data that does not shown all mandatory CDF fields for readability. The optional fields are shown in Table~\ref{tab:cdf-meta-optional}. 



Finally, if provided, skeletal tracking data should be accompanied by a skeletal hierarchy in the meta data under "meta/landmarks\_nodes".  It should adhere to the format shown in Figure~\ref{fig:skeletal-hierarchy} which shows a simplified hierarchy of a standardized T-pose. The format is borrowed from the glTF2.0 file format ~\cite{gltf2_specification}. The "name" field specifies the joint under consideration, it follows a snake\_case formatting with the "right", "left" indicators as a suffix. The "rotation" field is a quaternion defining the rotation of the joint in a T-pose, this can generally be left as [0.0, 0.0, 0.0, 1.0], but should be adjusted for joints that do not make right angles with their parent node (e.g. "thumb\_right" or "eye\_left"). Translation should follow the "Y-up" approach such that $x$, $y$ and $z$ components specify the orientation in relation to their parent node. For example "hip" has a child-node "spine" (as defined by the landmark node indices in "children") and there for [0, 1, 0] indicates the spine goes up (1) in the Y-axis and does not deviate in the $x$ or $z$ direction compared to the hip.

\begin{center}
\begin{minipage}{0.8\linewidth}
\begin{lstlisting}
{
    "match" : {
        "id" : "74e6661c",
        "status" : {
            "is_neutral": false,
            "has_extratime" : true,
            "has_shootout" : true
        },
        "result": {
            "final": {"home": 2, "away": 2, "winning_team_id": "3f029694"},
            "first_half": {"home": 1, "away": 0},
            "second_half": {"home": 2, "away": 2},
            "first_extratime": {"home": 3, "away": 2},
            "second_extratime": {"home": 3, "away": 3},
            "shootout": {"home": 4, "away": 5}
        }
    }    
}
\end{lstlisting}
\captionof{figure}{Simplified example of a partial JSON Object that depicts the structure of the match sheet data. In particular, it describes the format of the results in the this data source.}
\label{fig:match-data-json}
\end{minipage}
\end{center}

\begin{center}
\begin{minipage}{0.8\linewidth}
\begin{lstlisting}
{
    "meta": {
        "tracking": {
            "fps" :25,
            "version": "0.1.0",
            "name": "company_a"
        },
        "event": {
            "version": "0.1.0",
            "name": "company_b"
        },
        "meta": {
            "version": "0.0.1",
            "name": "company_b"
        },
        "cdf": {
            "version": "0.0.1",
            "name": "company_a"
        }
    },   
    "competition": {
        "id": "c155537a"
    },
    "season": {
        "id": "a3410mad"
    },
    "match": {
        "id": "74e6661c",
        "kickoff_time": "2024-08-29T14:00:00",
        "periods": [{
            "period": "first_half",
            "play_direction": "left_right"
        }, {
            "period": "second_half",
            "play_direction": "right_left"
        }],
        "whistles": [{
                "type": "first_half", "sub_type": "start", "time": "2024-08-29T14:00:00"
            },
            ...
        ]
    },
    "stadium": {
        "id": "0c82e46a", "pitch_length": 105.0, "pitch_width": 68.0
    },
    "teams": {
        "home": {
            "id": "11c4abab2025", "name": "FC Dagstuhl",
            "players": [
                    {
                        "id": "61a2b9ba", "team_id": "11c4abab2025",
                        "jersey_number": 46, "is_starter": true,
                        "name": "Gottfried Wilhelm Leibniz",
                        "first_name": "Gottfried", "last_name": "Leibniz",
                    } ...
            ]
        },
        "away": {
            "id": "3f029694", "name": "Dagstuhl United",
            "players": [
                    {
                        "id": "3f029694", "team_id": "3f029694"
                        "name": "Player Z", "jersey_number": 4, "is_starter": false,
                        ...
                    } ...
            ]
        }
    },    
    "referees": [{"id": "56c3b419"}]
}
\end{lstlisting}
\captionof{figure}{Example of a non-nested JSON ($.json$) meta data file with stadium, player, and team information.}
\label{fig:meta-data}
\end{minipage}
\end{center}

\subsubsection{Event and Tracking Data File Format}

Event and tracking (both centroid and skeletal) are provided in JSON Lines where every line contains a single JSON Object. The line separator between objects is ``{\textbackslash}n''~\cite{jsonlines}. Skeletal (limb) tracking (if provided) should be delivered in a separate file because this avoids the overhead and complexity of having to process a much larger file for the many applications that only require centroid data. 

Figure~\ref{fig:event-data} illustrates the first two events of the first half.  Figures~\ref{fig:tracking-data} shows a simplified example for two frames of player center of body tracking data.  Note that we are only showing the mandatory CDF fields. These can be augmented with the optional fields (see Tables~\ref{tab:cdf-event-optional} and~\ref{tab:cdf-com-tracking-optional}). If skeletal (limb) tracking is provided, it should follow the same structure as the tracking data JSON Lines format. Instead of a players $x$ and $y$ coordinates, the provider would provide coordinates for each landmark (e.g.,  $left\_hip\_x$, $right\_hip\_y$, $right\_hip\_x$) as specified by Table~\ref{tab:cdf-limb-tracking-required}.
The coordinate system should be the same used in the center of body tracking data, e.g., $x$, $y$, $z$ and are relative to the pitch. The field $is\_visible$ indicates if the part was actually measured, or inferred via an algorithm due to being hidden. Providers have flexibility over which body parts they track. This should be communicated beforehand.





\begin{center}
\begin{minipage}{0.8\linewidth}
\begin{lstlisting}
{
    "match": {"id": "74e6661c"},
    "meta": {"is_synced": true},
    "event" {
        "id": "7230efg1",
        "time": "2024-08-29T14:01:03",
        "period": "first_half",
        "type": "pass",
        "sub_type": "kick_off",
        "outcome": true,
        "outcome_detailed": "successful",
        "player_id": "c83323fb",
        "team_id: "3f029694",
        "body_part": "right_foot",
        "receiver_id": "da8e7c48",
        "receiver_time": "2024-08-29T14:01:05",
        "related_event_ids": ["32afa686"]
    }
}
{
    "match": {"id": "74e6661c"},
    "meta": {"is_synced": true},
    "event" {
        "id": "32afa686",
        "time": "2024-08-29T14:01:08",
        "period": "first_half",
        "type": "pass",
        "sub_type": null,
        "outcome": false,
        "outcome_detailed": "out_of_play",
        "player_id": "da8e7c48",
        "team_id: "3f029694",
        "body_part": "right_foot",
        "receiver_id": null,
        "receiver_time": null,
        "related_event_ids": null
    }
}
\end{lstlisting}
\captionof{figure}{Example of JSON Lines ({$.jsonl$}) that covers the mandatory CDF fields for Event Data. The file describes two events which are expanded over multiple lines for readability. }
\label{fig:event-data}
\end{minipage}
\end{center}

\begin{center}
\begin{minipage}{0.8\linewidth}
\begin{lstlisting}
{
    "frame_id": 0,
    "period": "first_half",
    "match": {"id": "74e6661c"},
    "teams": {
        "home": {
            "id": "11c4abab2025", 
            "players": [{"id": "61a2b9ba", "x": 0.13, "y": 4.04},  
                ...
            ]
        },
        "away": {
            "id": "3f029694", 
            "players": [{"id": "3f029694",  "x": 0.76, "y": 13.37}, 
                ...
            ]
        }
    },
    "ball": { 
        "x": 0.01,
        "y": 23.10,
        "z": 0.33,
        ...
    }
}
{
    "frame_id": 1,
    "period": "first_half",
    "match": {"id": "74e6661c"},
    "teams": {
        "home": {
                ...
            ]
        },
        "away": {
                ...
            ]
        }
    },
    "ball": { 
        ...
    }
}
\end{lstlisting}
\captionof{figure}{Example of the  JSON Lines ({$.jsonl$}) for the Center of Body Tracking Data representation. Two frames are shown and they are expanded over multiple lines for readability. We omit the positions of all players for readability.}
\label{fig:tracking-data}
\end{minipage}
\end{center}

\begin{center}
\begin{minipage}{0.8\linewidth}
\begin{lstlisting}
{
    "frame_id": 0,
    "period": "first_half",
    "match": {
        "id": "74e6661c"
    },
    "teams": {
        "home": {
            "id": "11c4abab2025",
            "players": [
                {
                    "id": "61a2b9ba",
                    "landmarks": [
                        {
                            "index": 0,
                            "name": "nose",
                            "x": 0.0,
                            "y": 0.0,
                            "z": 0.0,
                            "children": [1, 2, 3, 4],
                            "is_visible": true
                        },
                        {
                            "index": 1,
                            "name": "left_eye",
                            "x": 0.0,
                            "y": 0.0,
                            "z": 0.0,
                            "children": [],
                            "is_visible": true
                        },
                        {
                            "index": 2,
                            "name": "right_eye",
                            "x": 0.0,
                            "y": 0.0,
                            "z": 0.0,
                            "children": [],
                            "is_visible": true
                        },
                        {
                            "index": 3,
                            "name": "left_ear",
                            "x": 0.0,
                            "y": 0.0,
                            "z": 0.0,
                            "children": [],
                            "is_visible": true
                        },
                        {
                            "index": 4,
                            "name": "right_ear",
                            "x": 0.0,
                            "y": 0.0,
                            "z": 0.0,
                            "children": [],
                            "is_visible": true
                        },
                        // ... More landmark_nodes
                    ]
                },
                // ... More players
            ]
        },
        "away": {
            ...
        }
    }
}
\end{lstlisting}
\captionof{figure}{Example of JSON Object containing the standardized skeletal hierarchy of the provider specific skeletal model. The coordinate system is the same used in \ref{p:coords}. }
\label{fig:skeletal-hierarchy}
\end{minipage}
\end{center}







\section{Conclusions}
This paper proposes a new uniform and standardized format, called the Common Data Format (CDF), for five sources of data that are commonly collected during football (soccer) matches: match sheet data, video footage, event data, tracking data, and match meta data. 
This effort was motivated by the specific challenges many different end-users face when working with these data, as presently significant variation exists across providers in terms of what is collected, which specifications are used, how the data is represented, and how the data is delivered. The CDF provides uniform, standardized and unambiguous schema. As it was conceived with analysis in mind, it contains the minimal amount of information needed to perform common analysis tasks. Understanding that demands will vary between user groups, it has been designed with a minimum of mandatory fields while allowing unlimited optional fields for those requiring. To this end, the CDF should be seen as a continually evolving format, that will require updating and refinement as new data types emerge or definitions evolve to become standardised. However, this initial version represents an important step forward in improving the current state of play for end-users,  proposing a concrete file format and delivery method that enable both live and post-match processing in a unified manner.

\section*{Acknowledgments}
JD received support from the Flemish Government under the “Onderzoeksprogramma Artifciële Intelligentie (AI) Vlaanderen” program.  

\newpage

\begin{appendices}
\section{CDF Naming Conventions}
\label{app-naming-conv}

When defining CDF a number of conventions are used:

\begin{itemize}
    \item The field names use British spelling (``s" instead of ``z", and ``ou" instead of ``o" etc.).
    \item All common data format names are in xx\_yy (lowercase split by underscore), that is, ``snake case"
    \item CDF naming follows the following convention:
        \begin{itemize}
            \item Metric or object as first element (e.g., time, pass, speed, angle, player, match.)
            \item description that adds context to the metric (e.g., tracking, event, id, name, number.)
        \end{itemize}
    \item Use ``jersey" instead of shirt
    \item Do not include the units in the variable name, but include them in the description

\end{itemize}

\section{Optional Data for Each CDF Data Source}
\label{app:optional}
Beyond the mandatory data, the CDF foresees the possibility to include optional data. If a provider elects to include optional data included in the following tables, they must adhere to the CDF names and values. 

We would like to emphasize that providers are free to include other optional information beyond what is discussed in this appendix as they see fit. If they do so, it must respect the conventions outlined in Subsection~\ref{cdf:represent_conventions} and Appendix~\ref{app-naming-conv}. 

\subsection{Optional Match Sheet Data}
\label{app:official-match-data}


Table~\ref{tab:cdf-matchdata-optional} lists optional match data. The fields in this table primary focus on providing additional information about the players as well as the coaches and referee. 

\begin{table}[H]
\centering
\caption{Description of the optional Match Sheet Data CDF Fields. Any description ending with a $^\star$ is repeated once per person (e.g., player, referee, coach). For players, only those named to the match day squad are considered. Note that \{i\} is used as an index for the items in a list.}
\label{tab:cdf-matchdata-optional}
\begin{adjustbox}{max width=\textwidth}
\begin{tabular}{lp{2cm}p{7cm}l}
\hline
Root & Field & Description & Type \\ \hline
teams/\{home\text{\textbar}away\} & short\_name & Short name of the home or away team & String \\
teams/\{home\text{\textbar}away\} & formation & Formation label of the home or away team (e.g. "4-4-2"). See~\ref{app:positions} for more information. & String \\
teams/\{home\text{\textbar}away\}/players/\{i\} & alternative\_id & Additional identifier(s) of the player$^\star$ & String \\
teams/\{home\text{\textbar}away\}/players/\{i\} & maiden\_name & Maiden name$^\star$. For example, "Smith" for Sophia Wilson. & String \\
teams/\{home\text{\textbar}away\}/players/\{i\} & short\_name & Short name$^\star$. For example "Mohamed Salah Hamed Mahrous Ghaly" as "Mo Salah" (or "Mohamed Salah") or "Givanildo Vieira de Sousa" as "Hulk". & String \\
teams/\{home\text{\textbar}away\}/players/\{i\} & position\_group & Position group acronym given according to the CDF-compatible groups given in Appendix \ref{app:positions}$^\star$ & String \\
teams/\{home\text{\textbar}away\}/players/\{i\} & position & Position label acronym per player given according to the CDF-compatible labels given in Appendix \ref{app:positions}$^\star$ & String \\
teams/\{home\text{\textbar}away\}/players/\{i\} & is\_captain & Whether the player is a captain (true) or not (false)$^\star$ & Boolean \\
teams/\{home\text{\textbar}away\}/players/\{i\} & date\_of\_birth & A player's date of birth in YYYY-MM-DD format$^\star$ & String \\
teams/\{home\text{\textbar}away\}/players/\{i\} & height & Height of a player in cm$^\star$ & Integer \\
teams/\{home\text{\textbar}away\}/players/\{i\} & foot & A player's dominant foot, which can take the values left, right or both$^\star$ & String \\
teams/\{home\text{\textbar}away\}/coaches/\{i\} & id & Unique identifier for a coach$^\star$ & String \\
teams/\{home\text{\textbar}away\}/coaches/\{i\} & first\_name & First name$^\star$ & String \\
teams/\{home\text{\textbar}away\}/coaches/\{i\} & last\_name & Last name$^\star$ & String \\
teams/\{home\text{\textbar}away\}/coaches/\{i\} & short\_name & Short name, a combination of First name and Last name (e.g. Jane Doe, not J. Doe)$^\star$ & String \\
referees/\{i\} & official\_type & The type of referee (e.g. video\_assistant\_referee, main\_referee, assistant\_referee or fourth\_official & String \\
referees/\{i\} & first\_name & First name$^\star$ & String \\
referees/\{i\} & last\_name & Last name$^\star$ & String \\
referees/\{i\} & short\_name & Short name$^\star$ & String \\
\end{tabular}
\end{adjustbox}
\end{table}

\FloatBarrier 

\subsection{Optional Video Data}
Table~\ref{tab:cdf-video-optional} lists optional video data. 
\begin{table}[H]
\centering
\caption{Description of the Optional Video Footage CDF fields.}
\label{tab:cdf-video-optional}
\begin{adjustbox}{max width=\textwidth}
\begin{tabular}{lp{1.5cm}p{10cm}l}
\hline
Root & Field & Description & Type \\ \hline
recording/camera & x & x location of the camera in relation to the pitch center (m)  & Float \\
recording/camera & y & y location of the camera in relation to the pitch center (m)  & Float \\
recording/camera & z & z location of the camera in relation to the pitch center (m)  & Float \\
\end{tabular}
\end{adjustbox}
\end{table}

\newpage
\subsection{Optional Event Data}
\label{app:event_data}



Table~\ref{tab:cdf-event-optional} describes options fields that may be included for event data in the CDF. These primarily focus on information that is useful when working with synchronize event and tracking data. Moreover, they include some metrics such as xG that are commonly of interest. 

\begin{table}[H]
\centering
\caption{Description of the optional Event CDF fields. If included, these fields are recorded once per event.}
\label{tab:cdf-event-optional}
\begin{adjustbox}{max width=\textwidth}
\begin{tabular}{lp{3cm}p{7cm}l}
\hline
Root & Field & Description & Type \\ \hline
tracking & frame\_id & Frame identifier from the tracking data related to the event at (x, y) as described in ~\ref{tab:cdf-event-required} & Integer \\
tracking & frame\_id\_end & Frame identifier from the tracking data related to the event at (x\_end, y\_end) as described in ~\ref{tab:cdf-event-required} & Integer \\
tracking/player & x & x location of the player committing the action according to the tracking data. For example, player committing a pass, or making a tackle (m) & Number \\
tracking/player & y & y location of the player committing the action according to the tracking data. For example, player committing a pass, or making a tackle (m) & Number \\
event & match\_clock & The match clock as a string in format "MM:SS.mm" (i.e. minutes, seconds, milliseconds) (e.g. "92:01.04"). The clock resets to "45:00.00", "90:00.00" or "105:00.00" at the start of "second\_half", "first\_half\_extratime", "second\_half\_extratime", respectively. & String \\
event/metrics & xg & Calculated xG value between [0,1] & Number \\
event/metrics & post\_shot\_xg & Calculated post-shot xG value between [0,1] & Number\\
event/metrics & xpass & Calculated expected pass value between [0,1] & Number \\
event/metrics & epv &  Expected possession value between [-1,1] & Number\\
event/metrics & packing\_traditional & Number of opposing players that have been outplayed by a pass according to the traditional packing approach. Using this approach, a player is outplayed when the player is closer to their own goal than the ball at the time of the pass and further away from the goal than the ball at time of reception. & Integer \\
event/metrics & packing\_horizontal & Number of opposing players that have been outplayed by a pass according to the horizontal packing approach. Using this approach, a player is outplayed when their horizontal line is crossed by the ball during the pass, i.e., between the time of the pass and the time of reception. & Integer \\
event/var & reviewed & Was this event reviewed by the video assistant referee (true) or not (false) & Boolean \\
event/var & upheld & Was the on-field ruling confirmed (true) or overturned (false) & Boolean \\
\end{tabular}
\end{adjustbox}
\end{table}

\newpage

\subsection{Optional Tracking Data}
\label{app:tracking_data}

Table~\ref{tab:cdf-com-tracking-optional} presents optional fields for center-of-mass. These include meta data, extra information about the game, tracking information for people other than players, and derived (physical) metrics for players.  

\begin{table}[H]
\centering
\caption{Description of optional center-of-mass tracking fields in the CDF that would be recorded once per frame. Any description ending with a $^\star$ is repeated once per person (e.g., player, referee) per frame. Note that \{i\} is used as an index for the items in a list.}
\label{tab:cdf-com-tracking-optional}
\begin{adjustbox}{max width=\textwidth}
\begin{tabular}{lp{2.5cm}p{7cm}l}
\hline
Root & Field & Description & Type \\ \hline
& ball\_status & Indicates if the ball is either in play (true) or out of play (false) & Boolean \\
& ball\_poss\_team\_id & Unique identifier of the team that currently possesses the ball & String \\
& ball\_poss\_status & Contextual description of the level of control (e.g., controlled, contested, ...) & String \\
vendor & \{event\text{\textbar}tracking\} & Vendor name for tracking or event data & String \\
teams/\{home\text{\textbar}away\} & name & Name of the away team & String \\
teams/\{home\text{\textbar}away\} & jersey\_colour & Jersey colour of the team as a hexadecimal color code (e.g. \#545B62 for {\textcolor[HTML]{545b62}{\rule{1em}{1em}}} or \#FFC107 for {\textcolor[HTML]{FFC107}{\rule{1em}{1em}}}). & String \\
teams/\{home\text{\textbar}away\} & formation & Formation label of the away team (e.g. "4-4-2"). See~\ref{app:positions} for more information. & String \\
teams/\{home\text{\textbar}away\}/players/\{i\} & lat & latitude of the player's position if the data is collected by GNSS or GPS $^\star$& Float \\
teams/\{home\text{\textbar}away\}/players/\{i\} & long &longitude of the player's position if the data is collected by GNSS or GPS$^\star$& Float \\
teams/\{home\text{\textbar}away\}/players/\{i\} & z & z location of the player on the pitch (m)$^\star$. See Paragraph~\ref{p:coords}. & Float \\
teams/\{home\text{\textbar}away\}/players/\{i\} & dist & Distance covered by the player/ball in the current frame (m)$^\star$ & Float \\
teams/\{home\text{\textbar}away\}/players/\{i\} & acc & Acceleration of the player/ball (m/s\textsuperscript{2})$^\star$ & Float \\
teams/\{home\text{\textbar}away\}/players/\{i\} & vel & Speed of the player/ball (m/s)$^\star$ & Float \\
teams/\{home\text{\textbar}away\}/players/\{i\} & is\_visible & Denotes if a player was visible (in view of the camera) (true) or not (false)$^\star$ & Boolean \\

teams/\{home\text{\textbar}away\}/players/\{i\} & minutes\_played & Exact minutes and seconds per player & Float \\

referees/\{i\} & id & Unique identifier for a referee$^\star$ & String \\
referees/\{i\} & x & x location of the referee on the pitch (m)$^\star$ & Float \\
referees/\{i\} & y & y location of the referee on the pitch (m)$^\star$ & Float \\
referees/\{i\} & z & z location of the referee on the pitch (m)$^\star$ & Float \\

\end{tabular}
\end{adjustbox}
\end{table}


\newpage

\subsection{Optional Meta-Data}
\label{app:meta-data}

Table~\ref{tab:cdf-meta-optional} continued in Table~\ref{tab:cdf-meta-optional-1} present optional fields for meta data. 

\FloatBarrier 

\begin{table}[H] 
\centering
\caption{Description of optional Match Meta CDF. Note that \{i\} is used as an index for the items in a list.}
\label{tab:cdf-meta-optional}
\begin{adjustbox}{max width=\textwidth}
\begin{tabular}{lp{3.4cm}p{7cm}l}
\hline
Root & Field & Description & Type \\ \hline
competition & name & Name of the competition & String \\
competition & format & Specifying the competition set-up such as league\_18 (i.e., league involving 18 teams), league\_20, knock\_out\_neutral, etc. & String \\
competition & age\_restriction & Does the competition have a restriction on player ages (U14, U15, U16, U17, U18, U19, U20, U21, U22, U23, None) & String \\
competition & type & Is it a youth, mens, womens, etc. competition & String \\
season & name & Name of the season & String \\
match & round & Round of the match (e.g., 1, 2, 3, final, semi\_final) & String \\
match & scheduled\_kickoff\_time & Scheduled kickoff time in UTC & String \\
match & local\_kickoff\_time & Scheduled kickoff time & String \\
match/misc & country & Country where the match was played & String \\
match/misc & city & City where the match was played & String \\
match/misc & precipitation & Precipitation in millimeters & Integer \\
match/misc & is\_open\_roof & Information whether the roof is closed (false) or open (false) & Boolean \\
stadium & name & Name of the stadium & String \\
stadium & turf & Information on the turf of the stadium (natural, natural\_reinforced, grass, clay, ...) & String \\

\end{tabular}
\end{adjustbox}
\end{table}

Continued on the next page.

\FloatBarrier 

\begin{table}[H] 
\centering
\caption{Description of optional Match Meta CDF continuation from Table~\ref{tab:cdf-meta-optional}. Note that \{i\} is used as an index for the items in a list. }
\label{tab:cdf-meta-optional-1}
\begin{adjustbox}{max width=\textwidth}
\begin{tabular}{lp{3.4cm}p{7cm}l}
\hline
Root & Field & Description & Type \\ \hline
periods/\{i\}  & type & Choose from "first\_half", "second\_half", "first\_half\_extratime", "second\_half\_extratime" or "shootout" & Float \\
periods/\{i\}  & time\_start & Timestamp (in UTC) of the start of the period & Float \\
periods/\{i\}  & time\_end & Timestamp (in UTC) of the end of the period & Float \\
periods/\{i\}  & frame\_id\_start & Frame identifier for the tracking data related to the start of the period & Float \\
periods/\{i\}  & frame\_id\_end & Frame identifier for the tracking data related to the end of the period & Float \\
periods/\{i\}  & left\_team\_id & Unique team identifier of the team playing on the left side of the pitch in this period. The CDF requires a standardized playing direction (see Section~\ref{sec:playing-direction}). The left\_team\_id and right\_team\_id should be the actual, non-standardized sides. & String \\
periods & right\_team\_id & Unique team identifier of the team playing on the right side of the pitch in this period.  & String \\
meta/system & tracking\_type & The type of tracking system (e.g., mobile, broadcast) that is used which can take on the values mobile, in\_stadium or broadcast & String \\
meta/system & tracking\_version & Version of the tracking system specifications & String \\
meta/system & event\_type & Event data collection method (e.g. on\_site, remote, speaker\_writer, ...) & String \\
meta/system & event\_version & Version of the event data collection specifications & String \\
meta/system & ball\_status\_type & Information on collection of ball status (algorithm, ...) & String \\
meta/system & ball\_possession\_type & Information on collection of ball possession (algorithm, ...) & String \\

\end{tabular}
\end{adjustbox}
\end{table}

\newpage

\section{Position Labels}
Formations as well as player positions are subjective. Multiple approaches have tried to objectify the discussions around formations and positions using tracking data \cite{Bialkowski2014,Bialkowski2015,Bauer2021_Formations_IEEE}. 
The take away learnings from these studies is that playing positions should be considered dynamic, and thus are constantly changing within a match. However, static formations defined for at least part of the matches are a relevant simplification used by experts.
In order to create a comparability among vendors, we map player positions from different sources to standardized positions.  Note that these positions describe pre-match assigned positions.

\begin{figure}[htbp]
  \centering
  \includegraphics[width=\textwidth]{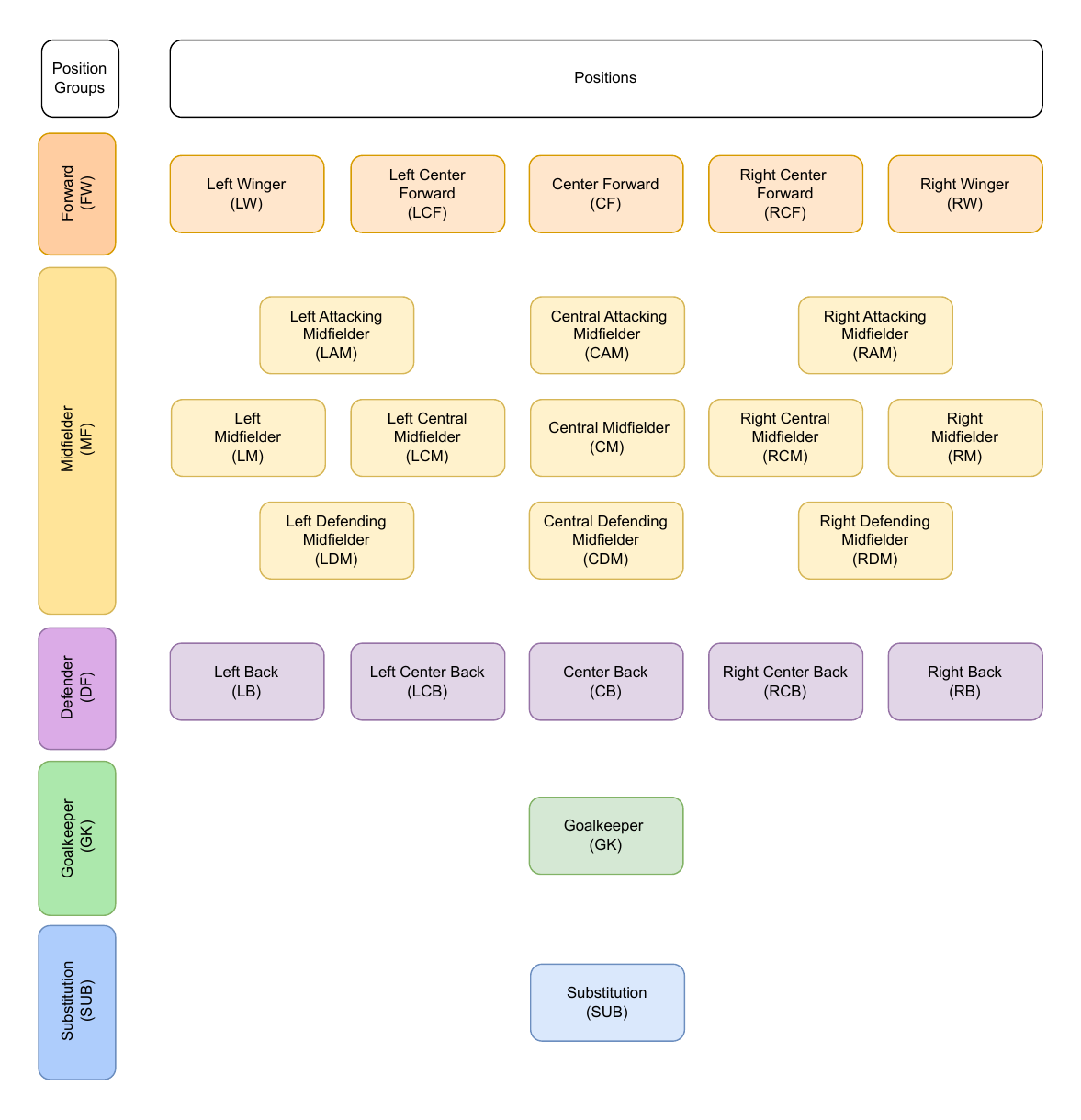}
  \caption{Position groups and associated position labels}
  \label{fig:positions}
\end{figure}

Position (`position`) and Position Group (`position\_group`) are both not mandatory under the CDF, but when provided positions and position groups should be taken from the diagram in \ref{fig:positions}.

These position labels have been identified from existing position labels provided by multiple providers (STS, Heimspiel, Noisefeed, SkillCorner, Impect and StatsPerform), but they have been filtered and aggregated with the aim of solely providing a spatial assignment and with the explicit aim to not be player roles. 

The labels that include Attacking and Defending are simply to distinguish how players from the same team (on average) relate to one-another in relative terms, and are not to be confused with a role assignment. They distinguish between midfielders that occupy spaces more forward or more backwards then their team mates. This should inform the reader why there is are no labels to Full Back, Wing Back, Inverted Wingback (Rightback \& Leftback), Second Striker or False 9 (Central Attacking Midfielder) or Box-to-Box Midfielder or  etc. 

We appreciate the distinction between positions and roles. We also understand that roles in particular can be fluid over the course of a match. Providers are welcome to define an extra optional field to denote a player's role. However, this must clearly be differentiated from the position groups defined in this appendix.

\subsection*{Position Labels Example}

Because of the uniqueness constraint there can always only be a maximum of one position label per team in play at any given moment. For example when a team is listed as playing a 4-4-1-1 the associated labels should be \textit{GK, LB, LCB, RCB, RB, LM, LCM, RCM, RM, CAM} and \textit{CF}, for a 4-4-2 the same would be true except that CAM and \textit{CF} are now replaced by \textit{LCF} and \textit{RCF} (or optionally \textit{LW} and \textit{RW}). 

Additionally, in case of a four or two player line, the `\textit{CF}`, `\textit{CM}` and/or `\textit{CB}` label(s) should not be used for that respective line. Instead, in case of a four player line the other four labels from that line should be used. And, in case of a two player line, the data provider can opt to use either pair of labels (ie. \textit{LM} and \textit{RM} or \textit{LCM} and \textit{RCM}). For a example in a 4-3-1-2, the defensive line should always be \textit{LB, LCB, RCB} and \textit{RB} and never \textit{LB, CB, CB, RB}, and the attacking line can be either \textit{LW} and \textit{RW} or \textit{LCF} and \textit{RCF}. When a team plays a three player line it should be \textit{LB, CB} and \textit{RB} and when they play a five player defensive line all five position labels should be used.

The above logic further dictates that a four or five player line can never consist of \textit{LAM, CAM, RAM}, or \textit{LDM, CDM} or \textit{RDM} and in this case labels should always be chosen to the five midfield labels (\textit{LM, LCM, CM, RCM} and \textit{RM}).

\subsection*{Formation Example}
The chosen position labels should inform the reported formation type and vice versa. This means that if the reported position labels are \textit{GK, LB, LCB, RCB, RB, LM, LCM, RCM, RM, CAM} and \textit{CF} the reported "formation" label should be 4-4-1-1 (a Sting-type with dashes in between each formation line). 

\subsection*{Section Highlights}

\begin{itemize}
  \item When providing position labels (`position`) there is a uniqueness constraint which means there can always only be one position label per team in play at any given moment.
  \item Any data provider is free to make additional custom labels that inform a players role, but by doing so they agree to also provide the `position` and `position\_group` labels following the CDF.
  \item The chosen position labels should match the chosen formation and vice versa.
  \item The formation should be denoted as a String with dashes between each line (e.g. "4-4-1-1").
\end{itemize}

\label{app:positions}

\vfill
\FloatBarrier 

\end{appendices}

\printbibliography
\end{document}